\documentclass[11pt,a4paper,twoside]{article}

\usepackage{graphics}

\usepackage{amssymb}

\newcommand{\reals}{\mathbb{R}}
\newcommand{\pc}{\mathbb{P}}

\begin{document}
\thispagestyle{plain}\enlargethispage{12pt}

\hspace{11cm} DAMTP-2002-66\\
\begin{center}
\textbf{\huge Born-Infeld Kinematics and\\[12pt] Correction to the Thomas Precession}\\[20pt]
{\large Frederic P Schuller\footnote{{\tt F.P.Schuller@damtp.cam.ac.uk}}}\\[10pt]
\textsl{DAMTP, University of Cambridge, Cambridge CB3 0WA}\\[10pt]

\begin{quote}
Dynamical symmetries of Born-Infeld theory associated with its maximal
field strength are encoded in a geometry on the tangent bundle of
spacetime manifolds. The resulting extension of
general relativity respecting a finite upper bound on accelerations is
put to use in the discussion of particle dynamics, first quantization,
and the derivation of a correction to the Thomas precession.\\[20pt]


\textsl{Journal Reference:} Physics Letters B \textbf{540},
119-124 (2002)
\end{quote}
\end{center}
\section{Introduction}
In classical mechanics, the study of phase space geometry yields
deep insights which would be hidden in a mere configuration space
formulation. In particular, there is no well-defined distinction between
coordinates and momenta, as these mix under symplectic transformations.\\
Geometric quantization aims at exploiting the symplectic structure of
phase space in order to understand the transition to quantum systems.\\
In contrast \cite{Born}, special and general relativity are formulated merely on
spacetime. This is of course likewise
true for all theories built on this framework, most notably quantum
field theory and string theory.\\[5pt]
% Born \cite{Born}, among others,
% regards a phase space formulation of general relativity as a necessary
% step towards a theory of quantum gravity.\\
Over the last two decades, there has been some interest in and
speculation on a finite upper bound on accelerations. This is mainly
due to the fact that although special relativity allows
arbitrarily high accelerations, upon quantization, a finite upper
bound enters through the back door \cite{Brandt2}.
This raises the question of whether to use kinematics respecting a
distinguished acceleration from start.
Indeed,
careful quantization of a particle with \textsl{dynamically} enforced submaximal
acceleration \cite{Nesterenko} nourishes the hope
that a finite upper bound on accelerations might positively influence
the convergence behaviour of field theoretic amplitudes. This
approach, however, makes an ad-hoc assumption of a maximal
acceleration and admittedly lacks a proper kinematical framework.\\[5pt]
The aim of this letter is to devise such kinematics, by kinematizing dynamical
symmetries of the Born-Infeld action on the velocity phase space of the
spacetime manifold, and to derive a correction to the Thomas
precession within this framework.

\section{Kinematization}
\label{Kinematization}
A particle of mass $m$ and electric charge $e$ minimally coupled to
Born-Infeld theory \cite{BornInfeld}
\begin{equation}\label{origBI}
  \mathcal{L}_\textrm{\small BI} = {\det}^{\frac{1}{2}}\left(g_{\mu\nu}+bF_{\mu\nu}\right)
\end{equation}
can at most experience an acceleration $\mathfrak{a} = e b^{-1}
m^{-1}$, as the maximal field strength is given by $b^{-1}$.\\
Denoting a point of the tangent bundle $TM$ of the $n$-dimensional spacetime manifold
$M$ by $X^m\equiv(\mathfrak{a} x^\mu, u^\mu)$, where $u^\mu$ is the
$n$-velocity of the particle, the Born-Infeld Lagrangian (\ref{origBI})
can be written
\begin{equation}\label{BIcommutator}
  \mathcal{L}_\textrm{\small BI} = {\det}^{\frac{1}{4}}\left([X^m,X^n]\right),
\end{equation}
if one assumes a $b^2$-suppressed coordinate
non-commutativity of spacetime in the presence of an electromagnetic field $F^{\mu\nu}$,
\begin{eqnarray}
  [x^\mu, x^\nu] &=& - i e^{-3} b^2 F^{\mu\nu},\\{}
  [x^\mu, p^\nu] &=& - i g^{\mu\nu},\\{}
  [p^\mu, p^\nu] &=& - i e F^{\mu\nu}.
\end{eqnarray}
Associated with the existence of a distinguished acceleration
$\mathfrak{a}$ there must be dynamical symmetries of Born-Infeld
theory. The form (\ref{BIcommutator}) suggests encoding these in
the geometry of the spacetime tangent bundle. We show that this
indeed results in a consistent kinematical framework that extends relativity
such as to respect a finite upper bound on accelerations.
Note that shifting the upper bound $\mathfrak{a}$ to infinity restores
coordinate commutativity. Hence spacetime
non-commutativity is a signature of a finite upper bound on
accelerations.

\section{Maximal Acceleration Geometry}
Consider the diagonal lift \cite{Yano1973} (here given in induced
tangent bundle coordinates $(x^\mu,u^\mu)$)
\begin{equation}
  g^D = \left(\begin{array}{cc} g_{ij} + g_{ts} {\Gamma^t}_i
  {\Gamma^s}_j
  & {\Gamma^t}_j g_{ti} \\ {\Gamma^t}_i g_{tj} & g_{ij}
  \end{array}\right)
\end{equation}
of the spacetime metric $g$ to the tangent bundle, where ${\Gamma^t}_i
\equiv u^a {{\Gamma_a}^t}_i$, and $\Gamma$ are the Christoffel symbols of $g$.
Requiring positivity
of the natural lift \cite{Yano1973} $x^* \equiv
\left(\mathfrak{a} x^\mu, \frac{dx^\mu}{d\tau}\right)$ of a timelike
spacetime curve $x$,
\begin{equation}
  g^D(dx^*,dx^*) > 0
\end{equation}
is equivalent to restricting the acceleration of the projection
$\pi(x^*)=x$ to values less than $\mathfrak{a}$. This was first
observed by Caianiello \cite{Caianiello1981} for flat spacetime.

\section{Complex Tangent Bundles?}
There were attempts \cite{Low,Brandt} to devise a maximal acceleration modification of
special and general relativity by equipping the tangent bundle with
the metric $g^D$ and an additional complex structure $F$, analogous to
the phase space structure of non-relativistic systems. However, invoking a
strong principle of equivalence, one must require both
structures to be covariantly constant,
\begin{equation}\label{spe}
  \nabla g^D = 0 \qquad \textrm{and} \qquad \nabla F = 0,
\end{equation}
where $\nabla$ is the Levi-Civit\`a connection with respect to $g^D$.
Then according to the Tachibana-Okumura theorem \cite{TO1962}, conditions
(\ref{spe}) are satisfied if, and only if, the base manifold $M$ is
flat.\\
Hence complex tangent bundles can never provide a theory of
gravity with finite upper bound on accelerations.\\

\section{Bimetric Tangent Bundles}
The Tachibana Okumura no-go theorem can be circumvented by using the
horizontal lift (here given in induced tangent bundle coordinates $(x^\mu,u^\mu)$)
\begin{equation}
  g^H = \left(\begin{array}{cc} u^a \partial_a g_{ij} & g_{ij} \\ g_{ji} & 0 \end{array}\right)
\end{equation}
of the spacetime metric instead of a complex structure. It can be shown \cite{Yano1973} that the horizontal lift
$\nabla^H$ of the
Levi-Civit\`a connection on spacetime is then the unique linear connection
on $TM$ satisfying
\begin{equation}\label{spe2}
  \nabla^H g^D = 0 \qquad \textrm{and} \qquad \nabla^H g^H = 0
\end{equation}
for arbitrary curvatures of the base manifold $M$. This, however,
comes at the cost of introducing torsion to the tangent bundle.
A tangent bundle curve $X: \reals \longrightarrow TM$ is called an
orbit if there exists a frame where $X = \pi(X)^*$. An orbit is called
an orbidesic if it is a geodesic with respect to some metric, or an
orbiparallel if it is an autoparallel with respect to some
connection.\\
The lifting and projection properties of spacetime geodesics and
tangent bundle orbidesics and orbiparallels are well known \cite{Yano1973}
and summarized in the diagram\\[10pt]

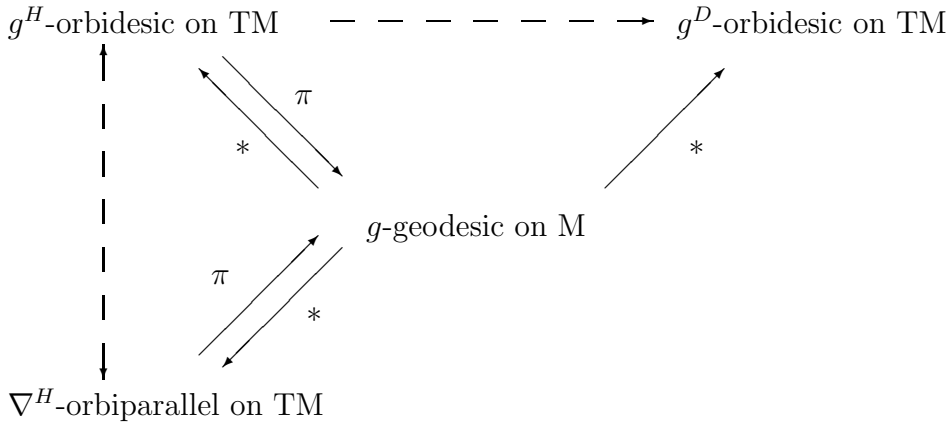
\setlength{\unitlength}{3947sp}%
\begingroup\makeatletter\ifx\SetFigFont\undefined%
\gdef\SetFigFont#1#2#3#4#5{%
  \reset@font\fontsize{#1}{#2pt}%
  \fontfamily{#3}\fontseries{#4}\fontshape{#5}%
  \selectfont}%
\fi\endgroup%
\begin{picture}(4512,2614)(526,-2444)
\thinlines
\put(1876,-136){\vector( 1,-1){750}}
\put(2476,-961){\vector(-1, 1){750}}
\put(4276,-961){\vector( 1, 1){750}}
\put(1726,-2011){\vector( 1, 1){750}}
\put(2626,-1336){\vector(-1,-1){750}}
\put(2551, 89){\line( 1, 0){150}}
\put(2851, 89){\line( 1, 0){150}}
\put(3151, 89){\line( 1, 0){150}}
\put(3451, 89){\line( 1, 0){150}}
\put(3751, 89){\line( 1, 0){150}}
\put(4051, 89){\line( 1, 0){150}}
\put(1126,-436){\line( 0,-1){150}}
\put(1126,-736){\line( 0,-1){150}}
\put(1126,-1636){\line( 0,-1){150}}
\put(1126,-286){\vector( 0, 1){225}}
\put(1126,-1036){\line( 0,-1){150}}
\put(1126,-1336){\line( 0,-1){150}}
\put(1126,-1936){\vector( 0,-1){225}}
\put(4351, 89){\vector( 1, 0){225}}
\put(526, 14){\makebox(0,0)[lb]{\smash{\SetFigFont{12}{14.4}{\familydefault}{\mddefault}{\updefault}
$g^H$-orbidesic on TM
}}}
\put(4726, 14){\makebox(0,0)[lb]{\smash{\SetFigFont{12}{14.4}{\familydefault}{\mddefault}{\updefault}
$g^D$-orbidesic on TM
}}}
\put(2776,-1261){\makebox(0,0)[lb]{\smash{\SetFigFont{12}{14.4}{\familydefault}{\mddefault}{\updefault}
$g$-geodesic on M
}}}
\put(1951,-811){\makebox(0,0)[lb]{\smash{\SetFigFont{12}{14.4}{\familydefault}{\mddefault}{\updefault}
*
}}}
\put(4801,-811){\makebox(0,0)[lb]{\smash{\SetFigFont{12}{14.4}{\familydefault}{\mddefault}{\updefault}
*
}}}
\put(2326,-436){\makebox(0,0)[lb]{\smash{\SetFigFont{12}{14.4}{\familydefault}{\mddefault}{\updefault}
$\pi$
}}}
\put(526,-2386){\makebox(0,0)[lb]{\smash{\SetFigFont{12}{14.4}{\familydefault}{\mddefault}{\updefault}
$\nabla^H$-orbiparallel on TM
}}}
\put(2401,-1861){\makebox(0,0)[lb]{\smash{\SetFigFont{12}{14.4}{\familydefault}{\mddefault}{\updefault}
*
}}}
\put(1801,-1561){\makebox(0,0)[lb]{\smash{\SetFigFont{12}{14.4}{\familydefault}{\mddefault}{\updefault}
$\pi$
}}}
\end{picture}\\

\noindent In particular, note that $g^H$-orbidesics coincide with
$\nabla^H$-orbiparallels, despite $\nabla^H$
having non-vanishing torsion \cite{Yano1973}. Further, one can show that all $g^H$-orbidesics
are $g^H$-null and $g^D$-positive.
\newpage
\section{Extended General Relativity}
The mathematical structure outlined above allows one to formulate physical
postulates for the kinematics of an extension of general relativity
respecting a finite upper bound on accelerations.
\begin{enumerate}
\item[I.] Submaximally accelerated particles are described by orbits
$X$ satisfying
  \begin{eqnarray}
    g^H(dX,dX) &=& 0, \label{gHnull}\\
    g^D(dX,dX) &>& 0  \label{gDpos}.
  \end{eqnarray}
\item[II.] In the absence of non-gravitational interaction, orbits of
particles are $g^H$-(null-)geodesics.
\item[III.] The physical time experienced by an observer with orbit $X$ is
measured by
\begin{equation}
  d\omega \equiv \left[g^D(dX,dX)\right]^{\frac{1}{2}}.
\end{equation}
\end{enumerate}
Note that $g^H$-orbidesics on $TM$ coincide with $g$-geodesics on
$M$, and are automatically $g^D$-positive. Hence, for unaccelerated
motion, the kinematics of standard general relativity are
recovered. This explains why for accelerations that are small compared
to the upper limit $\mathfrak{a}$, we only need one single metric $g$
on spacetime.\\
For non-$g^H$-geodesic motion, the metric $g^D$ assumes a non-trivial
r\^ole, as it restricts the deviation of orbits from $g^H$-orbidesics
to the $g^D$-positive ones.\\
Condition (\ref{gHnull}) enforces the orthogonality of the
$n$-velocity and $n$-acceleration of a particle in any frame. Thus
we now recognize that in standard general relativity already, the
horizontal lift $g^H$ is a \textsl{naturally induced} structure on the
associated tangent bundle.\\
Postulate III predicts a deviation from the relativistic physical
time for accelerated particles. From decay experiments \cite{Farley}
we get a lower bound for the maximal acceleration $\mathfrak{a}$ of
about $10^{19} m s^{-2}$. From high-precision measurements of the
Thomas precession, we get a better lower bound in section \ref{sec_thomas}.

\section{Field Equations}
It is conceptually inevitable to formulate the extended theory
entirely on the tangent bundle without recurring to spacetime
objects. The latter are viewed as derived concepts via the canonical
bundle projection. In this spirit we find \cite{bik_paper} the lifted Einstein field
equations
\begin{equation}\label{LEFE}
  (G^{ambn} - \frac{1}{2} G^{abmn}) R^V_{mn} = 8 \pi G T^D_{mn},
\end{equation}
where $R^V$ denotes the vertical lift \cite{Yano1973} of the spacetime Ricci tensor
$R$, and
\begin{equation}
  G^{abcd} \equiv g^{D_{ab}} g^{D_{cd}} + g^{H_{ab}}g^{H_{cd}}.
\end{equation}
The lifted equations (\ref{LEFE}) are equivalent to the Einstein field
equations on spacetime.

\section{Extended Special Relativity}
The case of flat Minkowski spacetime can be studied in a very concise
and illuminating way. The diagonal and horizontal lifts of the
Minkowski metric $\eta$ can be written
as
$$\eta^D = \eta \otimes 1 \quad \textrm{and} \quad \eta^H = \eta
  \otimes I,$$
with
\begin{equation}
  1 \equiv \left(\begin{array}{cc} 1 & 0 \\ 0 & 1
  \end{array}\right), \quad
  I \equiv \left(\begin{array}{cc} 0 & 1 \\ 1 & 0
  \end{array}\right).
\end{equation}
This motivates the study of the pseudo-complex
numbers
$$\pc \equiv \left\{a + I b | a,b \in \reals, I^2=+1\right\}.$$
As these build a commutative ring, pseudo-complex Lie algebras can be
defined \cite{Lang}. The pseudo-complexification $V_\pc$ of a real vectorspace
$V$ is a free module of pseudo-complex dimension $\dim_\reals V$, and we
have the isomorphism of $V_\pc \cong TV$ as real vectorspaces. If $V$ is
a representation space for a real Lie algebra $L$, then $L_\pc$ acts
naturally on $V_\pc$.\\
We are particularly interested in the pseudo-complexification of
Minkowski spacetime $(\reals^n,\eta)$ and the real Lorentz group
$SO_\reals(1,n-1)$. On the algebra level one gets
\begin{eqnarray}\label{pcso13}
  so_\pc(1,n-1) &=& \left<M^{\mu\nu}\right>_\pc\\
                &=& \left<M^{\mu\nu}, I M^{\mu\nu}\right>_\reals\nonumber
\end{eqnarray}
and obtains the connection component of the identity of the
pseudo-complex Lorentz group by exponentiation. Changing the basis in
(\ref{pcso13}) to $G^{\mu\nu}\equiv \frac{1}{2}(M^{\mu\nu} + I  M^{\mu\nu})$ and
$\bar G^{\mu\nu} \equiv \frac{1}{2}(M^{\mu\nu} - I  M^{\mu\nu})$ we get the
decomposition
$$
  so_\pc(1,m) = so_\reals(1,m) \oplus so_\reals(1,m).
$$
Thus, the representation theory in the pseudo-complex case can be
easily obtained from the real case.\\
Clearly, for a pseudo-complex $n$-vector $U^\mu \equiv u^\mu + I a^\mu$, the expression
\begin{eqnarray}
  \eta(U,U) &\equiv& U^\mu U^\nu \eta_{\mu\nu}\\
            &=& (u^\mu u_\mu + a^\mu a_\mu) + I (2 u^\mu a_\mu)\nonumber
\end{eqnarray}
is $SO_\pc(1,n-1)$-invariant, separately in the real and
pseudo-imaginary parts. This shows the isomorphism of
\begin{equation}
  \left(T\reals^n, \eta^D, \eta^H\right) \cong \left(\pc^n, \eta\right)
\end{equation}
as inner product spaces. Hence, in the flat case we can trade a
\textsl{bi}metric real vectorspace against a metric module. Essentially, special relativity with pseudo-complex coordinates is extended special relativity.\\
In particular, according to (\ref{pcso13}) all pseudo-complex Lorentz
transformations can be composed of the standard boost and rotation
transformations with pure real and pure pseudo-imaginary arguments.
For the interpretation of the pseudo-boosts, it is instructive to
consider the orbit $X$ induced by a spacetime curve describing a
submaximally accelerated hyperbolic motion. Using the
Lorentz-invariance of the theory, we can always arrange for this
motion to be in 1-direction in a rest frame at coordinate
time zero. As $d\omega$ is an $SO_\pc(1,3)$-invariant, we may study the
action of a pseudo-boost on the covariant velocity $U \equiv
\frac{dX}{d\omega} \equiv \tilde u + I \tilde a$. For hyperbolic
motion of spacetime curvature $g\equiv \mathfrak{a}\tanh(\alpha)$, the projections $\pi_{01}$ and
$\pi_{10}$ of $U$ into the
$\tilde u^0 - \tilde a^1$ and $\tilde u^1 - \tilde a^0$ planes,
respectively,  are straight lines through the origin of hyperbolic angle
$\tanh^{-1}\left(\frac{g}{\mathfrak{a}}\right)$:
\begin{eqnarray}
  \tilde a^0 &=& \frac{g}{\mathfrak{a}} \tilde u^1,\\
  \tilde a^1 &=& \frac{g}{\mathfrak{a}} \tilde u^0.
\end{eqnarray}
A pseudo-boost with boost parameter $I\beta$ can be easily seen to
hyperbolically rotate both of these lines by an angle $\beta$ within their
respective planes. Hence, the $\pi_{01}$ and $\pi_{10}$ projections of
the transformed curve coincide with the projections for a U-curve
corresponding to a hyperbolic motion with spacetime curvature
$\mathfrak{a}\tanh(\alpha+\beta)$. Carefully counting degrees of
freedom, one checks that the transformation of the two projections
already determines the transformation of the whole U-curve. Hence, the
pseudo-boosts are transformations to relatively uniformly
accelerated frames, respecting the maximal acceleration
$\mathfrak{a}$.\\
A pseudo-complex Lorentz tranformation of an orbit $X$ clearly induces
a transformation of the spacetime projection $\pi(X)$. For not purely
real transformation parameters, however, these transformations cannot be
understood as maps $M \longrightarrow M$, simply because the
components projected out by $\pi: TM \longrightarrow M$ contribute to the transformation.\\
In other words, spacetime events fail to be well-defined under
non-real Lorentz tranformations. Hence, we observe the breakdown of
the \textsl{classical} spacetime particle concept when changing to an
accelerated frame. This anticipates the Unruh effect \cite{Unruh} on a
classical level already.

\section{Dynamics}
Prior approaches \cite{Nestsecond} to maximal acceleration dynamics enforce
the finite upper bound on accelerations \textsl{dynamically}, i.e. by
modified Lagrangians, but still in the kinematical framework of special
or general relativity. The prototypical example is the massive
particle studied by Nesterenko et al. \cite{Nestaction},
\begin{equation}
  L = \sqrt{\ddot x^\mu \ddot x_\mu - \mathfrak{a}^2} \sqrt{\dot x^\mu
  \dot x_\mu} dt^2.
\end{equation}
This dynamical enforcement inevitably results in Lagrangians containing second order
derivatives, being inconvenient for technical and conceptual reasons,
especially in the transition to quantum theory.\\
In contrast, pseudo-complexification of merely relativistic
Lagrangians, e.g. for the massive relativistic particle to
\begin{equation}
 L_\pc = \sqrt{\dot X^\mu \dot X_\mu} dt,
\end{equation}
where
\begin{equation}\label{plift}
  X \equiv \mathfrak{a} x + I u,
\end{equation}
results in first order Lagrangians. From relation (\ref{plift}) this
might appear to be a mere notational trick. However, this is not the
case as $u$ and $x$ are independent degrees of freedom in extended
relativity, only linked by the pseudo-complex kinematics, in
particular the orthogonality condition \ref{gHnull}.\\
Studying the equations of motion for $L_\pc$  and making full use of
the pseudo-complex Lorentz symmetry, we indeed get a free particle
\cite{bik_paper}. A fully
$SO_\pc(1,3)$-invariant coupling term to Born-Infeld theory can be
constructed from the diagonal lift of the Kaluza-Klein metric from
spacetime to the tangent bundle \cite{bik_paper}. It turns out that
mere
pseudo-complexification of the standard minimal coupling term achieves
the same with much less labour.\\
Pseudo-complexification applies to relativistic spaces, symmetry algebras and
Lagrangians alike to give their extended relativistic
counterparts. This makes the pseudo-complex formalism so worthwile.

% \section{Quantization}
% Geometric quantization translates the symplectic structure of
% classical phase space into the commutator in the associated quantum
% theory. In extended relativity,
% the tangent bundle carries the pseudo-complex structure $\eta^H$
% instead of a complex structure. Using the Wigner \cite{Zachos} formalism, for
% example, to perform the transition to a first quantized theory, the
% pseudo-complex structure can be argued to give rise to anti-commutation
% relations\cite{bik_paper}. In the same sense as spinors are more fundamental than
% tensors, the anti-commutation relations are more fundamental as the
% commutation relations. Hence the absence of a complex structure in
% extended relativity does not obstruct at all the transition to
% quantum theory, but may shed a new light on the origin of
% anti-commutation relations.

\section{Correction to Thomas Precession\label{sec_thomas}}
Due to the non-commutativity of the boost generators
\begin{equation}
  [M^{0i},M^{0j}] = M^{ij}
\end{equation}
in special relativity,
an observer resting at the center of a uniform circular motion of
angular velocity $\omega$ and radius $R$ observes the Thomas precession \cite{Thomas}
\begin{equation}
  \frac{d\theta_{SR}}{dt} = (\gamma_{SR}-1)\omega
\end{equation}
of the spatial coordinate system of an orbiting observer, where
$\gamma_{SR} = (1+R^2\omega^2)^{1/2}$.
Exactly along the same lines, the non-commutativity of the
pseudo-boost generators
\begin{equation}
  [I M^{0i},I M^{0j}] = M^{ij}
\end{equation}
leads to an additional precession around the same axis, effecting a
total precession rate
\begin{equation}
 \frac{d\theta_{ESR}}{dt} = (\gamma_{ESR}-1)\omega,
\end{equation}
where $\gamma_{ESR} =
(1+R^2\omega^2+R^2\omega^4\mathfrak{a}^{-2})^{1/2}$. From
experimental data \cite{Newman}, obtained at
$R\omega^2=10^{18}ms^{-2}$, $\gamma_{SR}=1.2$, the ratio
$\frac{\gamma_{ESR}}{\gamma_{SR}}$ deviates from unity by less
than $5 \times 10^{-9}$. This yields a lower bound
\begin{equation}
  \mathfrak{a} \geq 10^{22} ms^{-2}
\end{equation}
and hence an upper bound on the Born-Infeld parameter
\begin{equation}
  b \leq 10^{-11} \frac{C}{N}.
\end{equation}
Thus high-precision measurements of the Thomas precession might be
able to discriminate between Maxwell and Born-Infeld electrodynamics in
the future.

\section*{Acknowledgements}
I would like to thank Gary Gibbons for very helpful discussions and
remarks on the material of this paper. I have also benefitted from
remarks by Paul Townsend and discussions with Sven Kerstan.
This work is funded by EPSRC and Studienstiftung des deutschen Volkes.

%\begin{noteinproof}
% A note added in proof, if there is one, should be the final text
% before the references.
%\end{noteinproof}

\end{document}